\documentclass[journal]{IEEEtran}
\usepackage[utf8]{inputenc}
\usepackage{amsmath}
\usepackage{amsthm}
\usepackage{multirow}
\usepackage{amssymb}
\usepackage{xcolor}
\usepackage{graphicx}
\usepackage{float}
\usepackage{bm}

\usepackage{cite}
\usepackage{subfigure}
\usepackage{caption}
\usepackage{comment}
\usepackage{epstopdf}
\usepackage{float}
\usepackage{hyperref}
\usepackage{balance}

\usepackage{amsfonts}
\usepackage{amsthm}
\usepackage{subcaption}
\usepackage{tikz}

\usepackage[noend]{algpseudocode}
\usepackage[ruled,norelsize,linesnumbered]{algorithm2e}
\usepackage{multicol}
\usepackage{circledsteps}

\algdef{SE}[DOWHILE]{Do}{doWhile}{\algorithmicdo}[1]{\algorithmicwhile\ #1}%
\makeatletter
\def\BState{\State\hskip-\ALG@thistlm}

\newlength\myindent
\theoremstyle{remark}
\newtheorem{remark}{Remark}

\newtheorem{definition}{Definition}
\makeatletter
\newcommand{\removelatexerror}{\let\@latex@error\@gobble}
\makeatother

\title{\Large\bf Model-free Vehicle Rollover Prevention: A Data-driven Predictive Control Approach}
\author{ Mohammad R.~Hajidavalloo,
Kaixiang~Zhang,
Vaibhav Srivastava,
Zhaojian~Li$^*$
\thanks{\quad This work was partially supported by National Science Foundation grant CMMI-2320698 and the Automotive Research Center (ARC) in accordance with Cooperative Agreement W56HZV-19-2-0001 U.S. Army DEVCOM Ground Vehicle Systems Center (GVSC) Warren, MI.}
\thanks{\quad  Mohammad Hajidavalloo, Kaixiang Zhang, and Zhaojian Li are with the Department of Mechanical Engineering, Michigan State University, East Lansing, MI 48824, USA.
        Email: {\tt\small \{hajidava,zhangk64,lizhaoj1\}@msu.edu}.}
\thanks{\quad  Vaibhav Srivastava is with the Department of Electrical and Computer Engineering, Michigan State University, East Lansing, MI 48824, USA.
        Email: {\tt\small vaibhav@msu.edu}.}

\thanks{$*$ Zhaojian Li is the corresponding author.}
        }%

\IEEEoverridecommandlockouts
\begin{document}
\maketitle
\begin{abstract}
Vehicle rollovers pose a significant safety risk and account for a disproportionately high number of fatalities in road accidents. This paper addresses the challenge of rollover prevention using Data-EnablEd Predictive Control (DeePC), a data-driven control strategy that directly leverages raw input-output  data to maintain vehicle stability without requiring explicit system modeling. To enhance computational efficiency, we employ a reduced-dimension DeePC that utilizes singular value decomposition-based dimension reduction to significantly lower computation complexity without compromising control performance. This optimization enables real-time application in scenarios with high-dimensional data, making the approach more practical for deployment in real-world vehicles. The proposed approach is validated through high-fidelity CarSim simulations in both sedan and utility truck scenarios, demonstrating its versatility and ability to maintain vehicle stability under challenging driving conditions. Comparative results with Linear Model Predictive Control (LMPC) highlight the superior performance of DeePC in preventing rollovers while preserving maneuverability. The findings suggest that DeePC offers a robust and adaptable solution for rollover prevention, capable of handling varying road and vehicle conditions.
\end{abstract}
\section{Introduction}
Vehicle rollovers remain a significant challenge in road safety, particularly for high-center-of-gravity (CG) vehicles such as sport utility vehicles (SUVs) and light trucks. Rollovers are associated with a disproportionate number of fatalities compared to their frequency, emphasizing the need for effective prevention strategies.
The National Highway Traffic Safety Administration (NHTSA)\cite{nhtsa,larish2013new} reports that vehicle rollovers constitute about 3\% of passenger-vehicle crashes each year. However, rollover accidents are responsible for 33\% of fatalities in all passenger-vehicle crashes.

Vehicle rollovers can be categorized into two types: tripped and untripped. Tripped rollovers typically occur when a vehicle collides with an object, such as a curb or obstacle, causing the rollover. In contrast, untripped rollovers, often induced by the driver, can happen during routine driving situations and pose a significant risk.
Examples include excessive speed during cornering, sudden obstacle avoidance, or aggressive lane-change maneuvers, where rollover results from the forces generated by the wheels during these actions.
However, such rollover incidents can be prevented by monitoring vehicle dynamics and applying timely control interventions. This highlights the need for advanced driver assistance systems that remain unnoticeable under normal driving conditions but activate as needed to help maintain vehicle stability during extreme maneuvers\cite{solmaz2007methodology,carlson2003optimal}. 

Vehicle electronic stability control (ESC) and rollover prevention systems have gained significant research interest in recent years \cite{cao2011editors,park2015integrated,de2012wheel,li2016three,chen2001differential}. Most ESC systems utilize active yaw moment control to modify steering characteristics. However, vehicles with a high center of gravity and a narrow track width, such as SUVs, remain prone to rollovers, even when ESC is employed to maintain lateral stability. To address this issue, various techniques have been developed for rollover prevention, including active differential braking \cite{solmaz2006methodology,schofield2006vehicle,yoon2007design,lee2013rollover,solmaz2008adaptive,qian2020rollover}, active suspension \cite{yoon2008unified,yim2010design}, semi-active suspension adjustments \cite{yu2008heavy}, and active steering angle corrections \cite{solmaz2007methodology,adireddy2011mpc}. Active braking reduces lateral acceleration and velocity by applying brakes to the outer wheels when rollover risk is detected. Along this direction, the work in \cite{yim2012design} developed a robust controller combining active suspension with differential braking, enhancing the system's resilience to variations in vehicle parameters such as CG height and speed. By addressing these variations explicitly, the controller improved rollover prevention performance across diverse operating conditions.
\cite{johansson2004untripped} demonstrated the application of a gain-scheduled linear quadratic (LQ) controller in conjunction with differential braking to manage wheel lift-off during extreme maneuvers. Their approach leveraged convex optimization for efficient control allocation, ensuring that braking forces were distributed optimally to maintain stability.
In \cite{li2016three}, a 3D dynamic control active braking framework is proposed using model predictive control (MPC) to integrate yaw stability and rollover prevention. By predicting future vehicle states from driver inputs and sensor data, MPC applies preemptive corrections, managing yaw and rollover risks simultaneously. This approach addresses challenges like brake actuator delays and balancing competing stability objectives, showcasing the complexity of vehicle dynamics.

Additionally, active or semi-active suspension systems have been leveraged to generate a counteracting roll moment to resist rollover. In this regard, \cite{yim2010design} proposed a controller for rollover prevention combining active suspension and an electronic stability program (ESP). The active suspension, designed using a linear quadratic static output feedback approach, reduces roll angle and suspension stroke by controlling suspension stroke and tire deflection. However, it introduces an oversteer tendency, impacting maneuverability. To address this, ESP was incorporated. In \cite{chokor2017rollover}, the proposed controller features a hierarchical architecture with two levels. The upper-level controller, based on Lyapunov theory, calculates a virtual control input representing an additive roll moment around the roll axis to ensure the vehicle's roll converges to its desired value. The lower-level controller translates this roll moment into effective active suspension forces at each corner of the vehicle. The results of J-turn and Fish-hook tests conducted at an initial speed of $130 km/h$ demonstrated the proposed controller's effectiveness in preventing rollover without reducing the vehicle's speed. Active steering directly alters tire forces or the vehicle's trajectory to prevent rollover.
\cite{solmaz2007methodology} developed a robust active steering controller to maintain a stable load transfer ratio (LTR), a key rollover indicator. The controller effectively manages high-CG vehicle dynamics during extreme maneuvers like sharp cornering and obstacle avoidance by adjusting the vehicle's trajectory to reduce rollover risk. Among these methods, active brake control has received the most attention due to its straightforward implementation within conventional ESC systems \cite{li2016three}.

Beyond braking and suspension systems, the literature also explored advanced sensing and control technologies. For instance, \cite{yim2011design} introduced a preview control strategy that leverages GPS and inertial measurement units (IMUs) to anticipate driver steering inputs. This approach allows the control system to prepare for potential rollover scenarios before they fully develop, enhancing the system's proactive capabilities. Similarly, a learning-augmented reference governor algorithm \cite{liu2020model} has been explored so
that violations of pre-specified constraints (LTR) are avoided through reference modification. This demonstrates the versatility of modern control methodologies in addressing complex vehicle dynamics.

Despite the aforementioned extensive research and promising advancements, several key challenges remain. In particular, most existing studies rely on accurate system models, which require extensive domain knowledge and labor-intensive parameter calibrations--both costly and time-consuming processes. In this paper, we tackle the rollover avoidance problem using Data-Enabled Predictive Control (DeePC). DeePC \cite{deepc1,Deepc} is a data-driven control framework that leverages historical data to predict and optimize the future behavior of a system, eliminating the need for an explicit mathematical model. Unlike traditional MPC, which relies on predefined dynamic models, DeePC directly utilizes raw input-output data from the system to make control decisions.
Since the rollover dynamics behavior is inherently nonlinear, one key advantage of this approach is that it can identify complex patterns in data that traditional rollover indices might miss, leading to more precise rollover risk assessments. In addition, DeePC allows for bypassing the requirement of detailed modeling, especially for our considered framework that involves complex dynamics and possess sensor noises and terrain/vehicle model uncertainties.
Due to the data-driven nature, this approach is versatile and can be readily adapted to various vehicle models and driving maneuvers, with the provision of sufficient and quality data. 

The major contributions of this paper are summarized as follows:
\begin{enumerate}
    \item \textit{Application of DeePC to Rollover Avoidance}: This work is the first to use DeepC for addressing the vehicle rollover prevention problem, circumventing the need of explicit dynamic models while managing the nonlinear and complex dynamics with safety constraints effectively.
    \item \textit{Dimension Reduction in DeePC for Enhanced Computational Efficiency}: The paper leverages the Singular Value Decomposition (SVD)-based dimension reduction technique to significantly lower computational complexity without compromising control performance, which is practically important for real-time application.

    \item \textit{Comprehensive Simulation and Benchmarking}: Extensive simulations on high-fidelity CarSim models (for both sedans and utility trucks) validate the effectiveness of the proposed DeePC approach. The results demonstrate superior rollover prevention and adaptability compared to the model-based MPC benchmark, even under challenging scenarios and varying road conditions.
\end{enumerate}

The remainder of this paper is organized as follows. Section \ref{sec:prelem} introduces the rollover prevention problem and the model-based approach to tackle the problem. Also, the fundamentals of data-enabled predictive control is discussed in this section. In Section \ref{sec:deepc_for_roll}, the details of DeePC setting for rollover prevention and reduced-dimension DeePC are presented. Section \ref{sec:implement} outlines the simulation setup and evaluates the performance of the proposed approach using high-fidelity CarSim models, highlighting the results for both sedans and utility trucks and in handling unexpected and challenging scenarios, such as driving on riverbed-like surfaces. Finally, Section \ref{sec:conclusion} concludes the paper.

\section{Problem Formulation and Preliminaries}\label{sec:prelem} 
In this section, we describe the problem of vehicle anti-rollover control, followed by traditional model-based control approaches. We will also present the preliminaries of DeePC.

\begin{figure}[!h]
  \centering
  \subfigure[Rear view.]{
    \includegraphics[width=1\columnwidth]{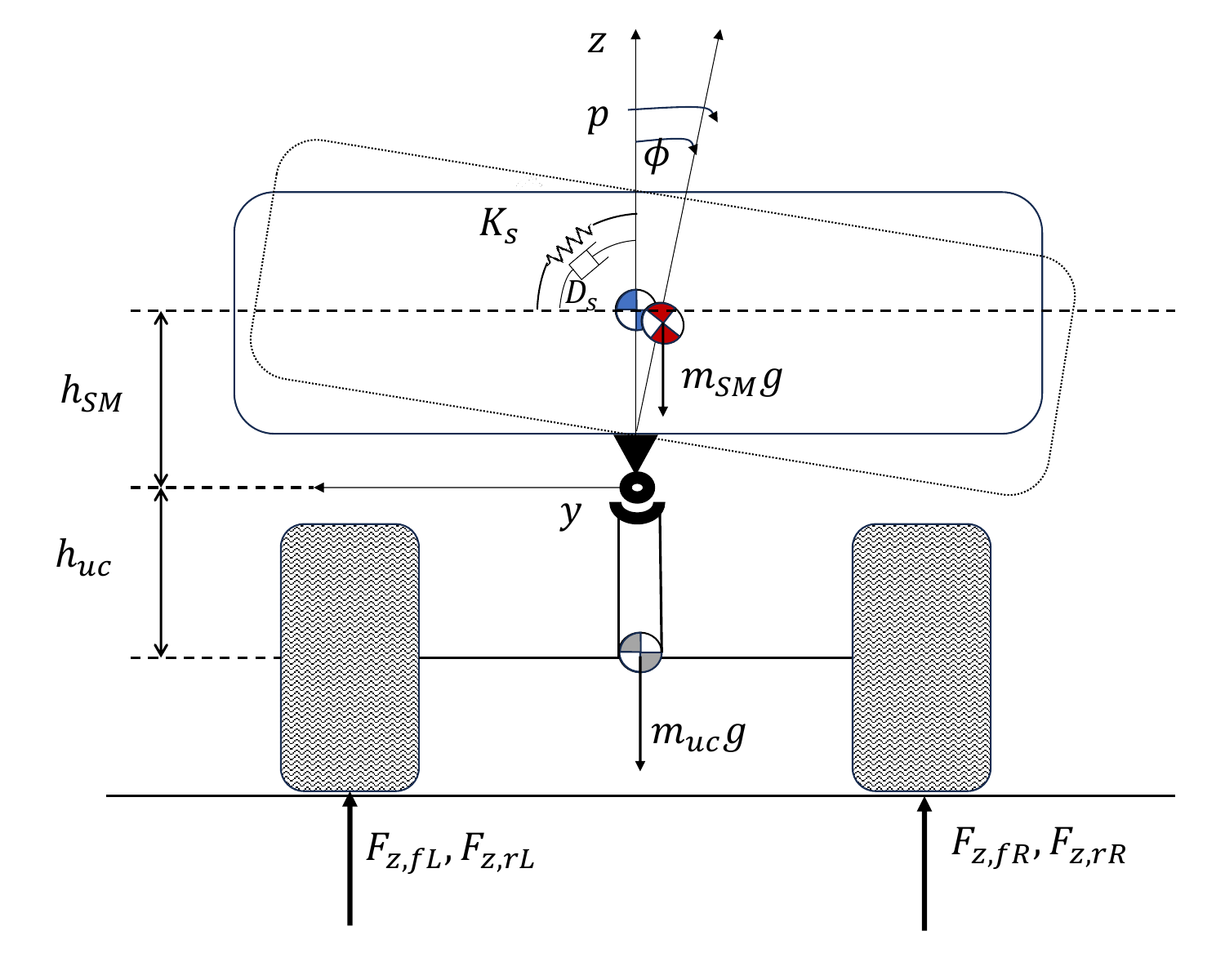}
    \label{fig:subfigA}
  }
  \subfigure[Top view.]{
    \includegraphics[width=1\columnwidth]{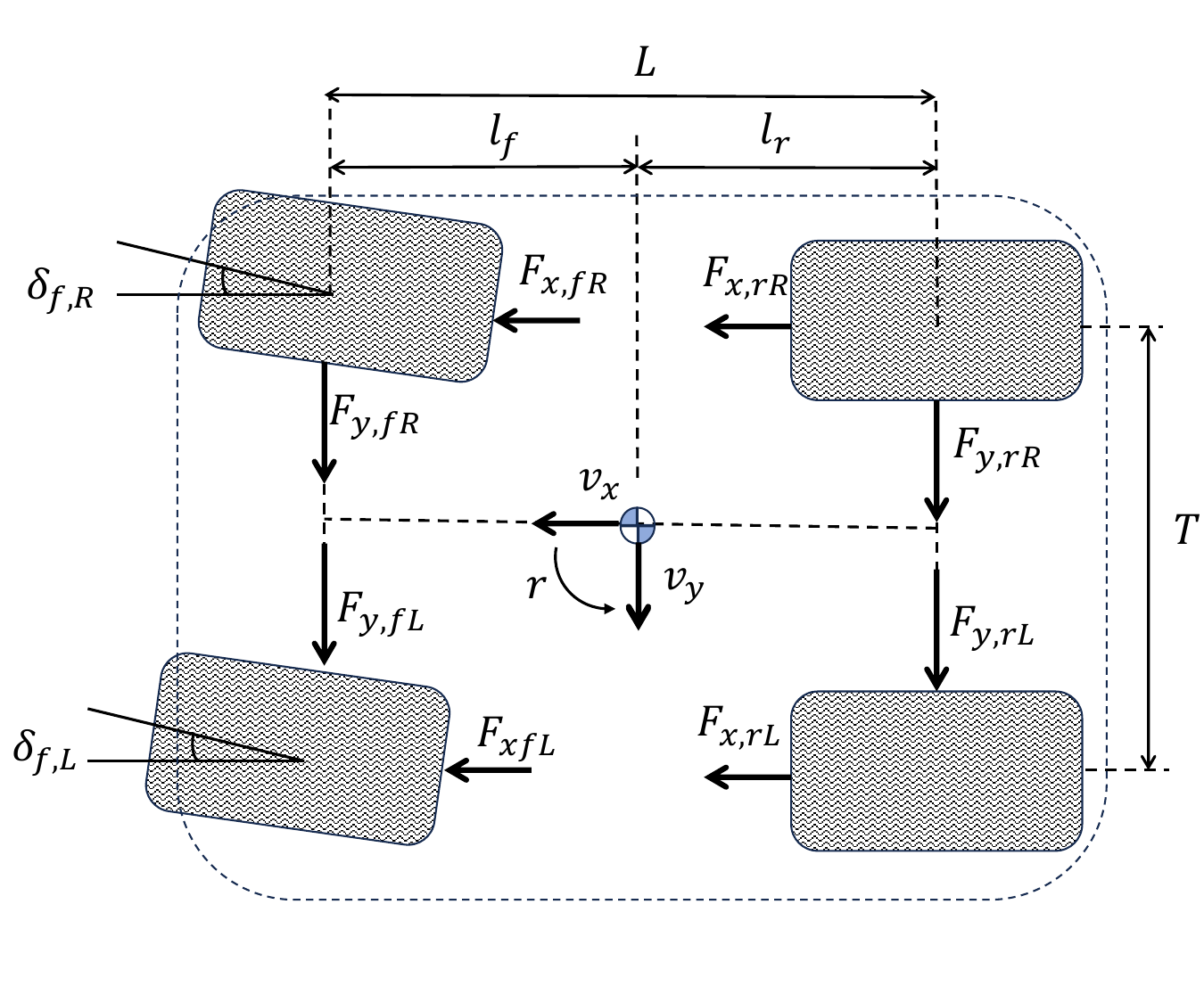}
    \label{fig:subfigB}
  }
  \caption{Schematics of vehicle force diagram during turning.}
  \label{fig:vehicle_force}
\end{figure}

\subsection{Vehicle rollover prevention}
Rollover is a critical safety concern in ground vehicles. The rollover prevention control problem focuses on designing control strategies that mitigate or prevent rollover incidents by actively regulating vehicle states, such as lateral load transfer, roll angle, and tire forces, through various actuation mechanisms. Vehicle rollover occurs primarily due to excessive lateral acceleration and abrupt maneuvers, which result in a significant shift of the normal tire forces, causing one or more wheels to lose contact with the ground \cite{Rajamani}. More specifically, considering a vehicle under steady state cornering with a lateral acceleration $\dot v_y$, let's denote the normal (vertical) load on the left tire by $F_{z,L}$ and on the right tire by $F_{z,R}$. The weight of the vehicle is
denoted by $mg$. The bank angle of the road is assumed to be negligible.  Taking moments about the right wheel contact point, and assuming the roll angle is small, we have 
\begin{equation}
    m\dot v_y+F_{z,L}l_w-mg\frac{l_w}{2} = 0 \Rightarrow F_{z,L} = \frac{mg\frac{l_w}{2}-m\dot v_y}{l_w},
\end{equation}
where $l_w$ is the track width of the vehicle. When the lateral acceleration of the vehicle is zero, the normal force on the inner wheels is 
$mg$. As the lateral
acceleration of the vehicle increases, the normal force decreases until it
becomes zero at a high enough lateral acceleration. This is the point at which
the inner wheels lift off the ground and this could be considered the
initiation of a rollover. 

Apart from steady-state conditions, the rollover phenomenon is commonly characterized by the LTR and roll moment stability criteria, which serve as key indicators of imminent rollover \cite{ungoren2004evaluation}. The LTR quantifies the distribution of vertical load across the vehicle's wheels and is given by
\begin{equation}
  \text{LTR} = \frac{F_{z,R}-F_{z,L}}{mg},  
\end{equation}
where $F_{z,R}$ and $F_{z,L}$ represent, respectively, the total vertical
force on the right-side tires and that on the left-side tires, and $mg$ is the vehicle weight. Rollover occurs when the LTR exceeds $1$ or drops below $-1$, indicating that all weight is concentrated on one side, leading to wheel liftoff.
Given the safety-critical nature of rollover prevention, modern control frameworks seek to optimize the trade-off between vehicle handling and stability while incorporating real-time sensor feedback and predictive modeling. The integration of autonomous and driver-assist technologies further necessitates robust, computationally efficient solutions that can respond to diverse driving scenarios and external disturbances. 

\subsection{System dynamics and nonlinear MPC-
based approach} \label{sec:nmpc-form}
Vehicle stability control approaches to rollover prevention typically involve active interventions using electronic stability control (ESC), differential braking, active suspension systems, and active steering \cite{abe2015vehicle}. These control strategies aim to redistribute forces within the vehicle’s dynamics to maintain stability, often leveraging model-based controls such as linear quadratic regulator control (LQR), MPC, or adaptive control techniques to ensure robust performance under varying operating conditions. Towards that end, we first introduce the nonlinear chassis and tire model necessary to formulate the model-based control approaches.
Following~\cite{solmaz2007methodology,zhou2009active,ulsoy2012automotive,bencatel2017reference} and under the assumptions that the sprung mass rotates around the center of mass of the undercarriage, the vehicle's inertia matrix is diagonal, and all wheels remain in contact with the ground, the nonlinear vehicle model can be derived as:
\begin{equation}
    \begin{aligned}
     &F_{x,T} = m(\dot v_x - v_yr) + m_{SM}h_{SM}p \cos \phi,
    \\&F_{y,T} = m(\dot{v_y} + v_xr) - m_{SM}h_{SM}(\dot p \cos \phi - p^2 \sin \phi),
    \\&L_T = -K_s(1 -\overline{\Delta k}_{ss}^2) \tan \phi - D_s(1 - \overline{\Delta d}_{ss}^2) p \cos \phi\\& \qquad \,\,\, - mg(\overline{\Delta k}_{ss} + \overline{\Delta d}_{ss}),
    \\&N_T = I_{zz}\dot{r},
    \\&\dot{p} = \frac{h_{SM}m_{SM}\left(\frac{F_{y,T}}{m} + \sin \phi (g + h_{SM}\frac{m_{uc}}{m}p^2) \right) + L_T}{I_{xx,SM} + h^2_{SM}m_{SM}\frac{m_{uc}}{m}\cos \phi},
    \end{aligned}
    \label{eq:veh-dyn}
\end{equation}
with most parameters illustrated in Fig.~\ref{fig:vehicle_force}. In~\eqref{eq:veh-dyn}, $v_x$ and $v_y$ are the vehicle's velocity components in the horizontal plane. $\phi$, $p$, and $r$ represent the roll angle, roll rate, and turn rate, respectively.
The forces and moments acting on the vehicle through the tires are represented by \(F_{x,T}\), \(F_{y,T}\), \(L_T\), and \(N_T\). The parameters \(m_{SM}, m_{uc},\) and \(m\) correspond to the sprung mass, undercarriage mass, and total vehicle mass, respectively, while \(I_{xx,SM}\) and \(I_{zz}\) denote the vehicle's moments of inertia along the $x$ and $z$ axes, respectively. The suspension system is characterized by roll stiffness \(K_s\) and damping coefficients \(D_s\), along with differential roll stiffness \(\overline{\Delta k}_{ss}\) and damping factors \(\overline{\Delta d}_{ss}\). The primary source of nonlinearities in the equations of motion arises from the dependence of tire forces on slip and ground contact force. The ground contact force, representing the vertical force at the tire contact patch, is unique for each tire and is denoted as \( F_{z(\cdot)} \), where \( (\cdot) = fR, fL, rR, rL \). The Magic Formula tire model \cite{ulsoy2012automotive,bencatel2017reference} is a widely used empirical model that describes the behavior of tires in various operating conditions. To determine the tire forces, the slip ratio \(\lambda\) and the tire slip angle \(\alpha\) are defined as
\begin{subequations}
\begin{align}
\lambda &= 
\begin{cases}
\frac{R_w\omega_w - u_w}{u_w}, & R_w\omega_w < u_w \\
\frac{R_w\omega_w - u_w}{R_w\omega_w}, & R_w\omega_w \geq u_w
\end{cases} \label{eq:lambda} \\
\alpha_f &= \delta_f - \tan^{-1}\left(\frac{v + l_fr}{u}\right), \label{eq:alpha_f} \\
\alpha_r &= \tan^{-1}\left(\frac{-v + l_rr}{u}\right), \label{eq:alpha_r}
\end{align}
\end{subequations}
where the tire slip ratio \(\lambda\) is defined as the normalized difference between the wheel hub speed \( u_w \) and the wheel's circumferential speed \( R_w \omega_w \), while the tire slip angle represents the angle between the tire's heading direction and the velocity vector at the contact patch. Equations \eqref{eq:alpha_f} and \eqref{eq:alpha_r} specify the tire slip angle for the front and rear wheels, respectively. When considering combined longitudinal and lateral slip, the Magic Formula is expressed as  \cite{bencatel2017reference}:
\begin{equation}
\begin{aligned}
\begin{bmatrix}
F_x^T\\
F_y^T
\end{bmatrix} &= F_P P(s_c, C_1, E_1)\hat{\mathbf{s}},\\
P(s_c, C_1, E_1) &= \sin(C_1\tan^{-1}[\frac{s_c}{C_1}(1-E_1)  \\ &+E_1\tan^{-1}\left(\frac{s_c}{C_1}\right)]),\\
s_c &= \frac{C_\alpha \|\mathbf{s}\|}{F_P}, \quad C_\alpha = c_1 mg\left(1-e^{-\frac{c_2 F_z}{mg}}\right), \\
c_1 &= \frac{B_1C_1D_1}{4\left(1-e^{-\frac{c_2}{4}}\right)}, \quad F_P = \frac{F_z 1.0527D_1}{1+\left(\frac{1.5 F_z}{mg}\right)^3},\\
\mathbf{s} &= \begin{bmatrix}
s_x\\
s_y
\end{bmatrix} = \begin{bmatrix}
\lambda\\
\tan \alpha
\end{bmatrix}, \quad \hat{\mathbf{s}} = \frac{\mathbf{s}}{\|\mathbf{s}\|}.
\end{aligned}\label{eq:magic}
\end{equation}
Here, \( F_x \) and \( F_y \) represent the forces along the tire's longitudinal and lateral axes, respectively (see Fig.~\ref{fig:vehicle_force}). \( C_\alpha \) denotes the cornering stiffness, while \( F_P \) refers to the peak horizontal (or slip) force. The total slip is represented by \( \hat{s} \), and the parameters \( B_1, C_1, D_1, E_1 \), and \( c_2 \) are tire-specific coefficients that vary based on tire characteristics and road conditions.
The complexity of the vehicle model, as evident from the equations, makes it challenging to accurately determine the real values of various parameters.
The complexity arises from the intricate relationships between system variables, external forces, and tire-road interactions, which complicates direct parameter identification and model calibration.

Given the vehicle dynamics equations \eqref{eq:veh-dyn}-\eqref{eq:magic}, consider a vehicle traveling from a starting point to an ending point along a road with sharp turns while maintaining a predefined high reference speed. The rollover avoidance problem is defined as determining the control inputs that prevent rollover, satisfy system constraints, and keep the vehicle's speed as close as possible to the reference speed with minor changes to reference steering, while ensuring both stability and maneuverability throughout the journey. The rollover avoidance problem described above can be formulated within a NMPC framework, as follows
\begin{equation}
\begin{aligned}
\min_{u} \sum_{t=0}^{N-1} &l(y(t), u(t) \\
\text{subject to } x(t+1) &= f(x(t), u(t)),  \\
y(t) &= h(x(t), u(t)), \\
u(t)& \in \mathcal{U},  y(t) \in \mathcal{Y},\\
x(0)& = x_0,
\end{aligned}
\end{equation}
where $l(u(t),y(t)) = \|u(t)-u_r(t)\|^{2}_R + \|y(t)-y_r(t)\|^{2}_Q$ is the stage cost, and $x = [v,p,r,\phi]^{T}$ and \(u = [\delta_f, v_x]^{T}\) are the state variable and control variables, respectively. For rollover prevention, various methods can be used to define constraints. One approach is to set constraints based on extreme roll angles where recovery is impossible.
In this paper, following \cite{solmaz2007methodology,solmaz2006methodology,bencatel2017reference} we consider a practical approach by enforcing rollover prevention constraints using the LTR. To ensure stability, we impose rollover avoidance constraints based on maintaining the LTR within safe bounds. In this paper, the LTR is taken as the control output ($y = \text{LTR}$), and the output and input constraints set are  defined as
\begin{equation}\label{eq:constraint}
\mathcal{Y} \in [-\text{LTR}_{lim},\text{LTR}_{lim}], \quad \mathcal{U} \in [-\delta_f, \delta_f] \times[v_{x,min},v_{x,max}],
\end{equation}
where $\text{LTR}_{lim} = 1$. In this regard, the functionals $f(\cdot, \cdot)$ and $h(\cdot, \cdot)$ are defined as the nonlinear relationships between the state and input pair and the next state and the output, respectively, as defined in equations \eqref{eq:veh-dyn}-\eqref{eq:magic}. 

The NMPC formulation above faces challenges associated with nonlinear vehicle modeling, such as difficulties in parameter identification and validation. Additionally, for varying road conditions, the nonlinear vehicle model requires continuous fine-tuning to maintain accuracy, and external factors such as road irregularities, varying bank angles, and abrupt driver inputs introduce uncertainties that are difficult to capture with a fixed mathematical model. 
Furthermore, a high-dimensional nonlinear constrained optimization problem needs to be solved at each time step, which is not practically viable due to the limited onboard computation and fast vehicle dynamics. In the next subsection, we will introduce a more practical linear MPC version that strikes a balance between performance and computation power. 


\subsection{Linear MPC-based rollover prevention control}
To address the computational complexity in the nonlinear MPC discussed above, we next introduce a linear MPC (LMPC) approach for rollover prevention, serving as a model-based control benchmark. The linear model representing the relationship between control inputs and output is given by:
\begin{equation}
    \begin{aligned}
    & x(t+1) = Ax(t)+Bu(t)\\
    &y(t) = Cx(t) + Du(t),
\label{eq:lin_model}
\end{aligned}
\end{equation}
which can be obtained through two primary approaches.  The first approach, similar to \cite{bencatel2017reference}, involves linearizing the nonlinear model presented in the previous section, followed by parameter identification to obtain the structured model based on observed data. This method leverages the given physics-based nonlinear model, linearizing it around a range of operating conditions.
The second approach  involves direct system identification without relying on a predefined structured model. Instead of deriving the linear model from a nonlinear formulation, a general state-space representation characterized by the system matrices 
$A,B,C,D$ is assumed. The elements of these matrices are then determined based on measured input-output data using system identification techniques such as subspace identification, least-squares estimation, or prediction error methods \cite{ljung1998system}. This approach is more useful when the system's physics are complex or unknown.
Similar to \cite{liu2020model}, in this work, we leverage the second approach for the derivation of the model matrices to perform the LMPC-based rollover prevention control.

A fourth-order model is chosen and system identification methods were used to identify the parameters. Similar to NMPC, here, the control inputs and outputs are $u = [\delta_f, v_x]^T$ and $y = \text{LTR}$. respectively. 
The parameters of the matrices are identified based on a CarSim model of the desired vehicle (sedan or truck) driving at a constant speed. 
Similar to the NMPC case, the rollover avoidance problem involves determining control inputs that prevent rollover, adhere to system constraints, and maintain vehicle stability and maneuverability while keeping speed and steering close to their reference values.
Consequently, the finite receding-horizon optimal control problem 
at each time step is defined as follows:
\begin{equation}\label{eq:MPC}
    \begin{aligned}
        \min_{u} \quad \sum_{t=0}^{N-1} l(u&(t), y(t)) \\
        \text{subject to:} \quad &\eqref{eq:lin_model},  \quad  x(0)  = x_0, \\ &u \in \mathcal{U}, \quad y \in \mathcal{Y},
    \end{aligned}
\end{equation}
where $l(u(t),y(t)) = \|u(t)-u_r(t)\|_R + \|y(t)-y_r(t)\|_Q$. Similar to NMPC, the control process is iterative. At each time step $t$, the first control action from the optimal sequence $u^*$ is executed, and the optimization is repeated for the next time interval.
LMPC relies on a linearized vehicle model, which may not accurately capture the highly nonlinear dynamics of rollover, especially during extreme maneuvers. This can lead to suboptimal or inaccurate control actions, especially when dealing with varying tire-road interactions, load transfer effects, and suspension dynamics. The need for precise parameter identification and adaptation to changing conditions increases the complexity of traditional model-based approaches. This further motivates the use of a data-driven framework like DeePC, which can adapt to different road and driving conditions without requiring explicit system identification. We next introduce the DeePC basics in the next subsection.

\subsection{Data Enabled Predictive Control (DeePC)} \label{sec: DeePC}
In this subsection, we introduce the basics of the DeePC framework that we will leverage for rollover prevention control in Section~III.

\subsubsection{Non-Parametric Representation of Linear Systems}
Consider the discrete-time linear time-invariant (LTI) system in \eqref{eq:lin_model} parameterized by $(A, B, C, D)$. 
Willems' fundamental lemma shows that it is possible to characterize \eqref{eq:lin_model}  using just a finite set of input/output data. Consider a input/output trajectory $u^d_{[0, T-1]}:=\mathrm{col}(u^d(0), u^d(1), \cdots, u^d(T-1) )$,  $y^d_{[0, T-1]}:=\mathrm{col}(y^d(0), y^d(1), \cdots, y^d(T-1) )$ of the system $\eqref{eq:lin_model}$ with length $T$. The corresponding Hankel matrices $\mathcal{H}_L(u^d_{[0,T-1]})$ and $\mathcal{H}_L(y^d_{[0,T-1]})$ take the form:

\begin{equation}
\left[\frac{\mathcal{H}_L(u_{[0, T-1]}^d)}{\mathcal{H}_L(y_{[0, T-1]}^d)}\right] \!:= \!
\left[\begin{array}{cccc}
u^d(0) & u^d(1) & \cdots & u^d(T\!-\!L) \\
\vdots & \vdots & & \vdots \\
u^d(L\!-\!1) & u^d(L) & \cdots & u^d(T\!-\!1) \\
\hline
y^d(0) & y^d(1) & \cdots & y^d(T\!-\!L) \\
\vdots & \vdots & & \vdots \\
y^d(L\!-\!1) & y^d(L) & \cdots & y^d(T\!-\!1)
\end{array}\right],
\label{eq:hankel_2}
\end{equation}
where each column represents a trajectory of length $L$ from system $\eqref{eq:lin_model}$. 
Before delving into the fundamental lemma, we first define the concept of persistent excitation.
\begin{definition}
	A sequence  $\omega_{\left[0,T-1\right]} := \mathrm{col} (\omega(0),\omega(1),\ldots,\omega(T-1)$ of length $T$ is said to be persistently exciting of order $L$ $(L \leq T)$ if its associated Hankel matrix
	\begin{equation}
		\mathcal{H}_{L}(\omega_{\left[0, T-1\right]}):=\begin{bmatrix}
			\omega(0) &\omega(1)& \cdots & \omega(T-L) \\
			\omega(1) &\omega(2)& \cdots & \omega(T-L+1)\\
			\vdots & \vdots & \ddots & \vdots \\
			\omega(L-1) &\omega(L)& \cdots & \omega(T-1)
		\end{bmatrix}
	\end{equation}
    has full row rank.
\end{definition}
When the input sequence exhibits persistent excitation, the column space of~\eqref{eq:hankel_2} provides a non-parametric description of $\eqref{eq:lin_model}$, as formalized in the following lemma.

\textit{Lemma 1 (Fundamental Lemma \cite{Depersis,willems2005note})}: Consider a controllable LTI system \eqref{eq:lin_model} and assume that the input sequence  $u_{[0, T-1]}^d$ is persistently exciting of order $n + L$, where $n$ is the dimension of the system state. Then, any length-$L$ sequence $(u_{[0,L-1]}, y_{[0,L-1]})$ is an input/output trajectory of~\eqref{eq:lin_model} if and only if we have
\begin{equation}
\left[\begin{array}{c}
u_{[0, L-1]} \\
y_{[0, L-1]}
\end{array}\right] = 
\left[\begin{array}{c}
\mathcal{H}_L(u_{[0, T-1]}^d) \\
\mathcal{H}_L(y_{[0, T-1]}^d)
\end{array}\right]g\
\label{eq:fund_lem}
\end{equation}
for some real vector $g \in \mathbb{R}^{T-L+1}$.

\subsubsection{Data-EnablEd Predictive Control}

Traditional control methods depend on an explicit model $(A, B, C, D)$ in $\eqref{eq:lin_model}$ for controller design. In contrast, DeePC \cite{Deepc}, \cite[Sec. 5.2]{markovsky2021behavioral} offers a non-parametric approach that eliminates the need for system identification and uses previously collected input/output data to construct a safe and optimal control policy. Specifically, DeePC leverages the historical data to predict system behavior by applying the fundamental lemma.

Consider integers $T_{\text{ini}}, N \in \mathbb{Z}$, and define $L = T_{\text{ini}} + N$. We choose a sufficiently 
long input sequence $u_{[0, T-1]}^d$ of length $T$, which is persistently 
exciting of order $n + L$. Let $y_{[0, T-1]}^d$ be the corresponding 
output sequence. The Hankel matrices $\mathcal{H}_L(u_{[0, T-1]}^d)$ 
and $\mathcal{H}_L(y_{[0, T-1]}^d)$ are partitioned into two segments: ``past data" of length 
$T_{\text{ini}}$ and ``future data" of length $N$:
\begin{equation}
\left[\begin{array}{c}
U_{\text{p}} \\
U_{\text{f}}
\end{array}\right] = \mathcal{H}_L(u_{[0, T-1]}^d), \quad
\left[\begin{array}{c}
Y_{\text{p}} \\
Y_{\text{f}}
\end{array}\right] = \mathcal{H}_L(y_{[0, T-1]}^d), \label{eq:hankel_1}
\end{equation}
where $U_{\text{P}}$ and $U_{\text{F}}$ correspond to $\mathcal{H}_L(u^d_{[0, T-1]})$'s first $T_{\text{ini}}$ block rows and last $N$ block rows, respectively (similarly for $Y_{\text{P}}$ and $Y_{\text{F}}$). Here, $u_{\text{ini}} = u_{[t-T_{\text{ini}},t-1]}$ denotes the input sequence over a past horizon of length $T_{\text{ini}}$, while $u = u_{[t,t+N-1]}$ represents the input sequence over a prediction horizon of length $N$ (with analogous definitions for $y_{\text{ini}}$ and $y$). At each time step $t$, DeePC solves the following constrained optimization problem:
\[\min_{g,u,y,\sigma_u,\sigma_y} \|y - y_r\|_Q^2 + \|u-u_r\|_R^2 + \lambda_u\|\sigma_u\|_2^2 + \lambda_y\|\sigma_y\|_2^2 + \lambda_g\|g\|_2^2\]
\begin{align}
\text{subject to} \quad & 
\begin{bmatrix}
U_P \\
U_F \\
Y_P \\
Y_F
\end{bmatrix} g = 
\begin{bmatrix}
u_{ini} \\
u \\
y_{ini} \\
y
\end{bmatrix} + 
\begin{bmatrix}
\sigma_u \\
0 \\
\sigma_y \\
0
\end{bmatrix}, \quad u \in \mathcal{U}, y \in \mathcal{Y},
\label{eq:deepc}
\end{align}
where $y_r = \mathrm{col}(y_r(t), y_r(t+1), \cdots, y_r(t+N-1))$ and $u_r = \mathrm{col}(u_r(t), u_r(t+1), \cdots, u_r(t+N-1))$ are the reference output and input trajectories, respectively, $Q \in \mathbb{S}_+^{pN}$, $R \in \mathbb{S}_+^{mN}$ are weighting matrices, $\mathcal{U}$, $\mathcal{Y}$ represent the input and output constraints, respectively, $\sigma_u \in \mathbb{R}^{mT_{ini}}$, $\sigma_y \in \mathbb{R}^{pT_{ini}}$ are auxiliary variables, and $\lambda_u \geq 0$, $\lambda_y \geq 0$, $\lambda_g \geq 0$ are regularization parameters.
DeePC solves \eqref{eq:deepc} in a receding horizon fashion. After computing the optimal sequence $u^* =\mathrm{col}(u_0^{*} \cdots u_{N-1}^{*})$, the first $l$ ($l < N$) elements $(u(t), \ldots, u(t+l-1)) = (u_0^*, \ldots, u_{l-1}^*)$ are applied to the system. Then, at time $t+l$, $(u_{ini}, y_{ini})$ is updated with the latest input/output data; see \cite{Deepc}, \cite{deepc1} for more details.

The DeePC algorithm was initially developed for deterministic linear time-invariant (LTI) systems. For nonlinear systems, a regularization strategy has been proposed to ensure robust performance \cite{Deepc,deepc1,coulson2019regularized,elokda2021data}, as reflected in \eqref{eq:deepc}. Specifically, when input/output data are collected from a nonlinear system, the subspace spanned by the resulting Hankel matrix may not align with the subspace of trajectories of the underlying system, even if the full-rank condition of the Hankel matrix is satisfied. Consequently, this misalignment can lead to poor prediction accuracy due to the ill-posed nature of the Hankel matrix constraint. To address this issue, a penalty term is introduced to account for the discrepancy—quantified by slack variables, $\sigma_u$, $\sigma_y$—between the estimated initial conditions $U_{P}g$, $Y_{P}g$ and the real-time buffered initial conditions $u_{ini}$, $y_{ini}$. This approach facilitates a least-squares estimation of the true initial condition, thereby mitigating prediction errors.

\begin{remark} [Dimension of $g$] \label{Remark:g}
For the sequence $u^d_{[0, T-1]}$ to satisfy persistent excitation, the Hankel matrix $\mathcal{H}_{n+L}(u_{[0,T-1]}^{d})$ must have at least as many columns as rows. This leads to the condition $T-(n+L)+1 \geq m(n+L)$, or equivalently, $T \geq (m+1)(n+L)-1$. Consequently, the dimension of $g$ in $\eqref{eq:deepc}$ has the following lower bound:
\begin{equation}
T-L+1\ge mL+(m+1)n
\end{equation}
where $L=T_{ini}+N$. Thus, a large $T$ results in a high-dimensional optimization variable $g$ in \eqref{eq:deepc}, which increases the computation burden and thus hinders the deployment of DeePC in resource-limited scenarios.
\end{remark} 

\begin{remark}[Practical selection of $T$] \label{Remark:T}
In \eqref{eq:deepc}, $T_{\text{ini}}$ should exceed the observability index to estimate the initial state $x_{\text{ini}}$ at time $t$ \cite[Lemma 4.1]{Deepc}, \cite[Lemma 1]{markovsky2021behavioral}. When the system model \eqref{eq:lin_model} is unknown, both this observability index (which is upper bounded by $n$) and the state dimension $n$ may be unknown. In practice, sufficient data points need to be collected to satisfy $T \geq (m+1)(n+L)-1$ where $L = T_{\text{ini}}+N$ \cite{elokda2021data,huang2021decentralized,wang2023deep,wang2023implementation}. This typically results in a very large $T$, making the optimization problem \eqref{eq:deepc} computationally challenging to solve efficiently due to its large scale.
\end{remark}

\section{DeePC for rollover prevention}
\label{sec:deepc_for_roll}
In this section, we present our main results on leveraging DeePC for data-driven rollover prevention. Note that the DeePC formulation in \eqref{eq:deepc} requires constructing the Hankel matrices \eqref{eq:hankel_2}, which consists of input and output data from the system. In our considered rollover prevention control problem, the input data comprises steering wheel angle and longitudinal speed, $u = [\delta_f, v_{x}]^T$, while the output data is the $y = \text{LTR}$.
In this regard, our approach can be seen as a data-driven counterpart of active steering control systems.
This choice is similar to the model-free approach incorporated by \cite{liu2020model}, which considered steering and LTR as control inputs and outputs, respectively, however, we extend this by adding target speed to enhance maneuverability during challenging driving scenarios.
For data collection, CarSim was utilized as the simulation platform for a scenario where the vehicle is traversing a road with multiple turns for a total of $30$ seconds.
The forward speed, which was set to $100 km/h$, and the steering wheel command, based on the Simple Driver Model provided in CarSim, are the recorded control inputs data where white Gaussian noise was introduced to the input in order to ensure persistent excitation of the system. The LTR data of the vehicle was recorded as the output of the system.
\begin{figure*}[!h]
    \centering
    \includegraphics[width=1 \linewidth]{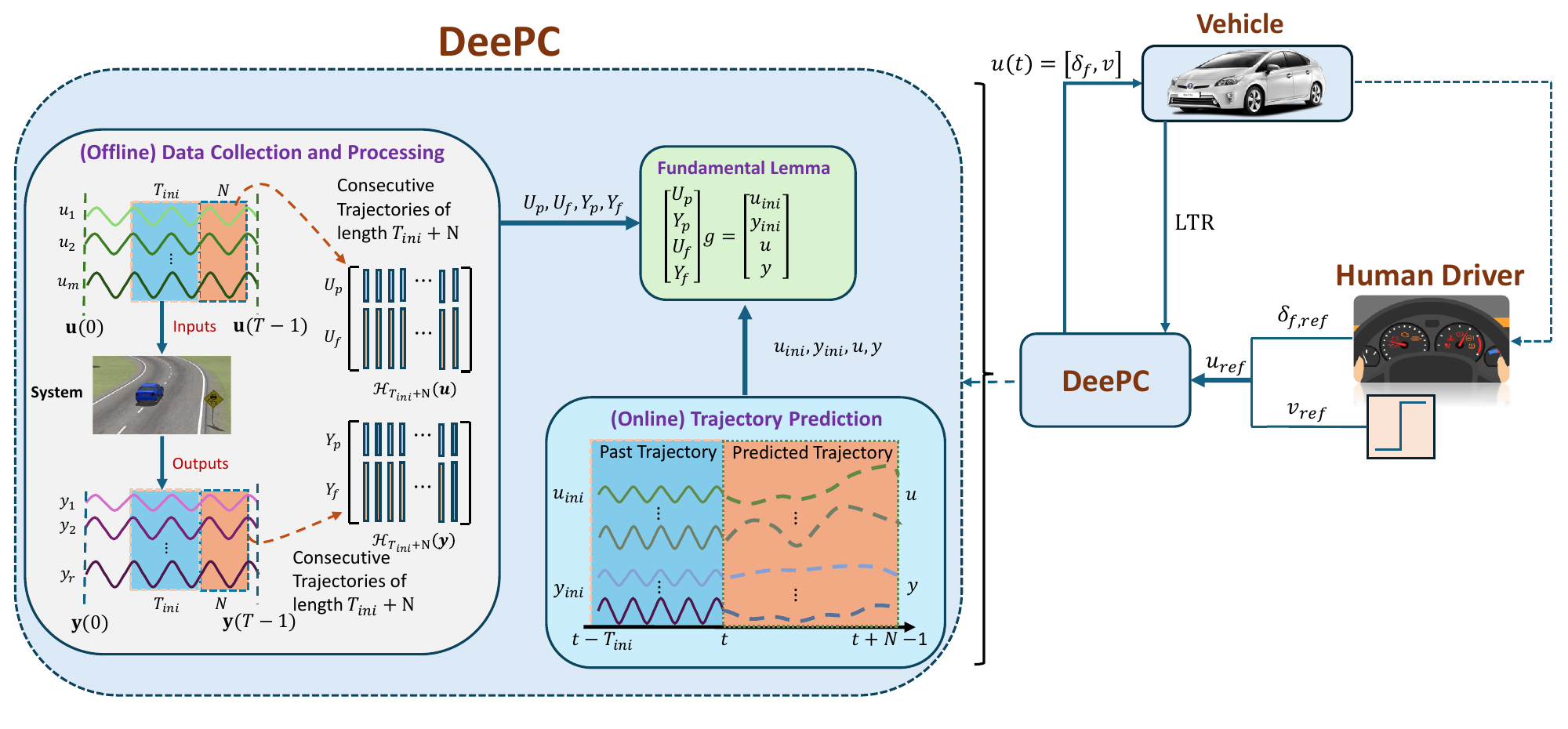}
    \caption{ Overview of the proposed framework using DeePC. }
    \label{fig:deepc-framework}
\end{figure*}
The input and output constraints set, $ \mathcal{U}, \mathcal{Y}$  in \eqref{eq:deepc} encompass permissible values for the vehicle's steering wheel angle, speed, and LTR, respectively. At this stage, all the components of \eqref{eq:deepc} have been identified, enabling the application of the DeePC framework to address the rollover prevention problem and we transform the optimization problem \eqref{eq:deepc} into standard quadratic programming.

\textcolor{black}{The overview of our main proposed control strategy framework, DeePC, is shown in Fig.~\ref{fig:deepc-framework}}
As shown in Fig.~\ref{fig:deepc-framework}, the primary function of the DeePC module is to modify control input commands originating from human-drivers, when needed, to prevent the rollover. More specifically, the DeePC module, utilizes historical input/output data to dynamically update the right-hand side of constraint \eqref{eq:deepc}. It also receives control references as input. The module's output is the updated control signal for the vehicle.

As discussed in Remarks~\ref{Remark:g} and \ref{Remark:T}, ensuring that the pro-collected data is sufficiently rich for non-parametric representation requires a large value of $T$. This typically results in the Hankel matrix in \eqref{eq:hankel_2} being short-fat, with its number of columns significantly exceeding its number of rows. This observation enables the use of a smaller data matrix to represent the input/output behavior of system \eqref{eq:lin_model}, which can be viewed as a \textit{reduced-dimension} version of the fundamental lemma. Based on our recent work~\cite{Zhang2023CSL}, we apply the singular-value decomposition (SVD) technique to reduce the dimension of the original Hankel matrix. This allows us to formulate a modified version of DeePC, effectively addressing the dimension issues outlined in Remarks~\ref{Remark:g} and \ref{Remark:T}.

The SVD of $\mathcal{H}_L(u^d_{[0,T-1]})$ and $\mathcal{H}_L(y^d_{[0,T-1]})$ can be expressed as
\begin{equation} \label{eq:svd}
\underbrace{\left[\begin{array}{c}
\mathcal{H}_L(u_{[0, T-1]}^d) \\
\mathcal{H}_L(y_{[0, T-1]}^d)
\end{array}\right]}_{\mathcal{H}_{L}} = \underbrace{[W_1 \quad W_2]}_{W} 
\underbrace{\left[\begin{array}{cc}
\Sigma_1 & 0 \\
0 & 0
\end{array}\right]}_{\Sigma}
\underbrace{[V_1 \quad V_2]^T}_{V^T},
\end{equation}
where $WW^T = W^TW = I_{(m+p)L}$ and $VV^T = V^TV = I_{T-L+1}$, and $\Sigma_1 \in \mathbb{R}^{q \times q}$ contains the top $q$ non-zero singular values. The reduced-dimension data matrix is then constructed as
\begin{align}
&\tilde{\mathcal{H}}_L = \mathcal{H}_L V_1 = W_1 \Sigma_1 \in \mathbb{R}^{(m+p)L \times q}.
\label{eq:svd2}
\end{align}
Note that the data matrix $\tilde{\mathcal{H}}_L$ has the same range space as $\mathcal{H}_{L}$, and thus $\tilde{\mathcal{H}}_L$ can be used to represent the input-output behavior of LTI system~\eqref{eq:lin_model}
.

After collecting the input/output data sequences $u_{[0, T-1]}^d$, 
$y_{[0, T-1]}^d$ and forming the Hankel matrices in \eqref{eq:hankel_1}, we 
can apply the SVD technique in~\eqref{eq:svd} and \eqref{eq:svd2} to compute a new 
data library $\bar{\mathcal{H}}_L$ and consequently achieve dimension reduction in DeePC.
Specifically, given $\bar{\mathcal{H}}_L$,  the DeePC problem in \eqref{eq:deepc} can be reformulated as
\[\min_{\bar{g},u,y,\sigma_u,\sigma_y} \|y - y_r\|_Q^2 + \|u-u_r\|_R^2 + \lambda_u\|\sigma_u\|_2^2 + \lambda_y\|\sigma_y\|_2^2 + \lambda_g\|\bar{g}\|_2^2\]

\begin{equation}    
\text{subject to} \quad \bar{\mathcal{H}}_L\bar{g} = 
\left[\begin{array}{c}
u_{\text{ini}} \\
u \\
y_{\text{ini}} \\
y
\end{array}\right] +
\left[\begin{array}{c}
\sigma_u \\
0 \\
\sigma_y \\
0
\end{array}\right], \, u \in \mathcal{U}, y \in \mathcal{Y}.
\label{eq:min_d_deepc}
\end{equation}
We refer to \eqref{eq:min_d_deepc} as the reduced-dimension version of DeePC, since the column dimension of $\bar{\mathcal{H}}_L$ is minimized while still guaranteeing the behavior representation of LTI systems. The optimization variable $g$ in \eqref{eq:deepc} has a dimension of $T-L+1$, whereas in \eqref{eq:min_d_deepc}, the dimension of $\bar{g}$ is reduced to $q$. This is expected to significantly reduce the computational complexity in optimization to enable real-time implementation.





\section{CarSim Simulations}\label{sec:implement}
 In this section, we conduct comprehensive simulations in CarSim 
to evaluate the efficacy of the proposed DeePC approaches. CarSim provides a realistic and high-fidelity simulation environment for simulating various vehicle behaviors under diverse road conditions, enabling an accurate assessment of control strategies. The simulations are designed to mimic real-world driving scenarios, including sharp turns, high-speed maneuvers, and different road conditions to test the effectiveness of the proposed control methods in preventing rollovers. Specifically, we  first investigate the effectiveness of the reduced-dimension DeePC in terms of performance and computation efficiency, as compared to LMPC and the standard DeePC. Then, the performance of the developed control strategies is thoroughly evaluated  with key performance metrics such as  LTR, vehicle speed, steering inputs, and stability to analyze and compare the capabilities of the DeePC and LMPC approaches. 

\subsection{Effectiveness of dimension reduction in DeePC}
\begin{figure}
    \centering
    \includegraphics[width=1\linewidth]{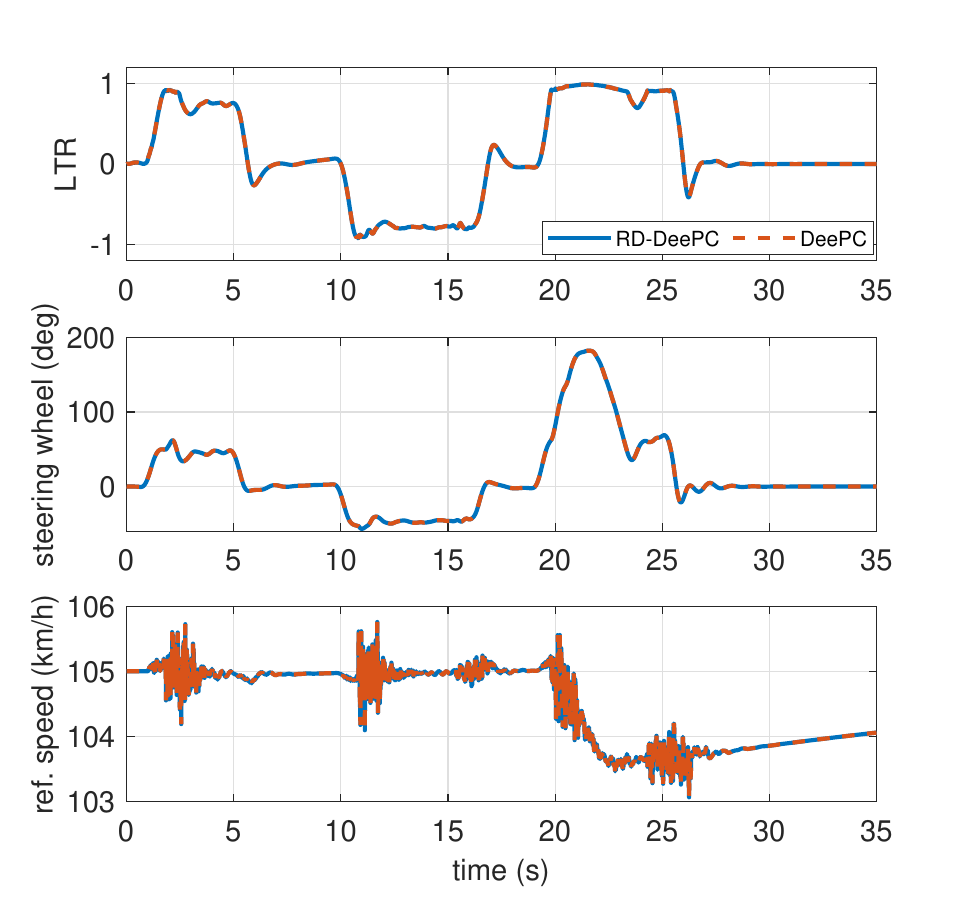}
    \caption{Comparison between DeePC and reduced-dimension DeePC}
    \label{fig:compare-deepc}
\end{figure}

The SVD-based dimension reduction technique is applied to the previously discussed DeePC formulation to demonstrate its ability to maintain optimal control performance while significantly reducing computational complexity. Specifically, input and output data are collected over a simulation period of $30$s with 0.01 sampling time, resulting in a Hankel matrix comprising $3001$ columns. The Hankel matrix is decomposed via SVD, and a truncated right-singular matrix $V_{[1:l]}$ is selected with $l = 600$ to construct the matrix $\tilde{\mathcal{H}}_L$. Fig.~\ref{fig:compare-deepc} illustrates the performance comparison between the original DeePC and the reduced-dimension DeePC. The close agreement in system inputs and outputs demonstrates that the proposed reduced-dimension DeePC achieves performance comparable to the original formulation. Additionally, the costs and computation times for both methods, as well as for LMPC, are summarized in Table~\ref{tab:min-deepc-vs-deepc}. While the costs of the two DeePC approaches are almost identical, the proposed efficient DeePC significantly reduces the time required for step-by-step optimization. Furthermore, in comparison with LMPC, the overall cost of DeePC and RD-DeePC, which is defined as $\|u(t)-u_r(t)\|_R^{2} + \|y(t)-y_r(t)\|_Q^{2}$ with the same weighting matrices $R$  and $Q$ for all strategies, is significantly lower.
Therefore, in the subsequent discussions and results, we adopt the reduced-dimension DeePC (RD-DeePC) as our approach. 

\begin{table}
\caption{DeePC performance metrics based on ${\mathcal{H}}_L$ and $\tilde{\mathcal{H}}_L$}
\begin{tabular}{ccc}
\hline
& Cost &  Time per Step [sec]  \\
\hline
DeePC, ${\mathcal{H}}_L \in \mathbb{R}^{600\times 3001}$  & $9.94$ & $1.446s$ \\
RD-DeePC, $\tilde{\mathcal{H}}_L \in \mathbb{R}^{600\times 600}$ & $9.95$ & 0.089s \\
LMPC & $74.13$& $0.004s$\\

\hline
\label{tab:min-deepc-vs-deepc}
\end{tabular}
\end{table}

\subsection{Sedan vehicle}
\begin{figure}
    \centering
    \includegraphics[width=0.9 \linewidth]{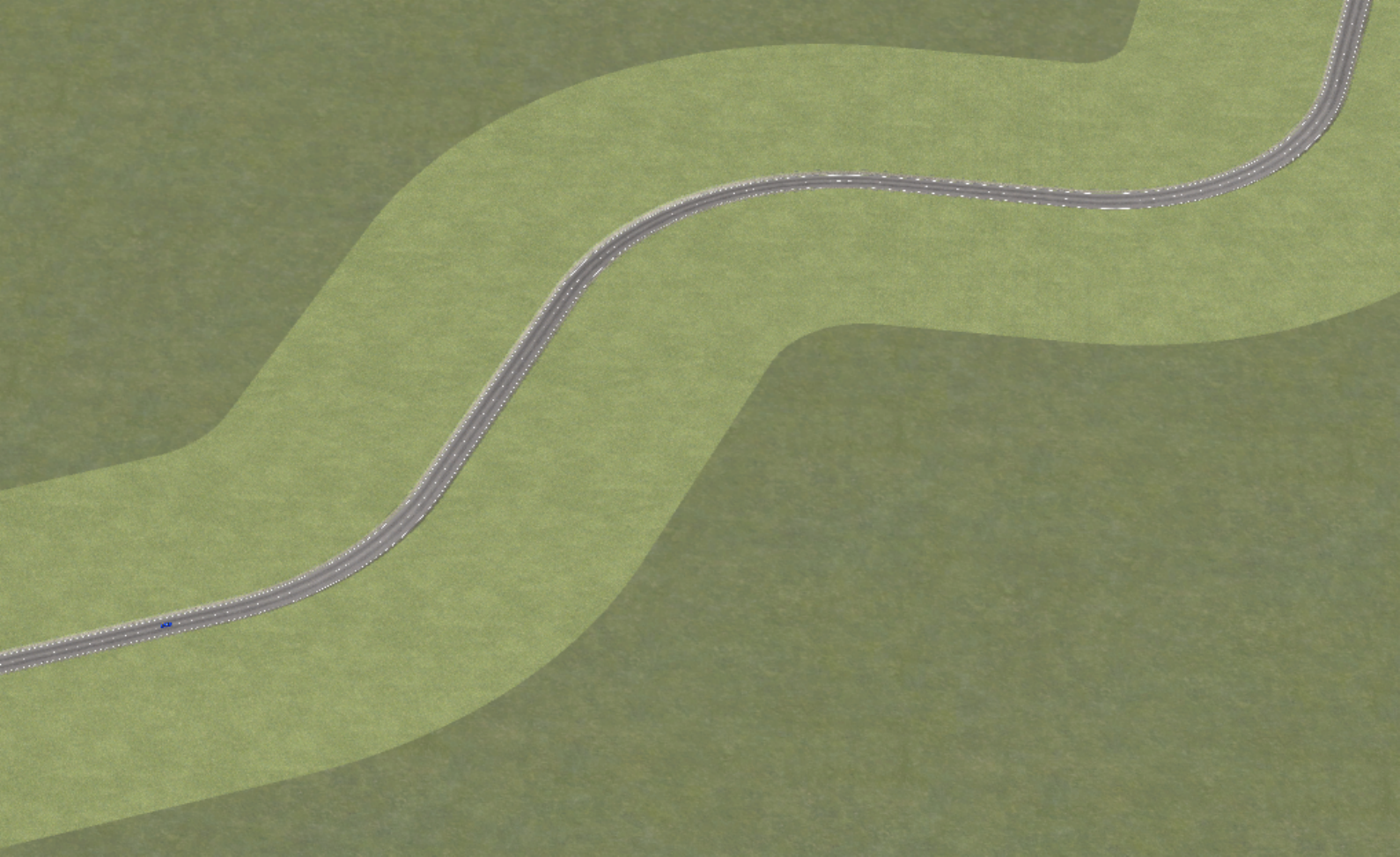}
    \caption{ Road geometry used in CarSim.}
    \label{fig:road_geo}
\end{figure}

\subsubsection{Simulation setting}

\begin{figure*}
    \centering
    \includegraphics[width=0.8 \linewidth]{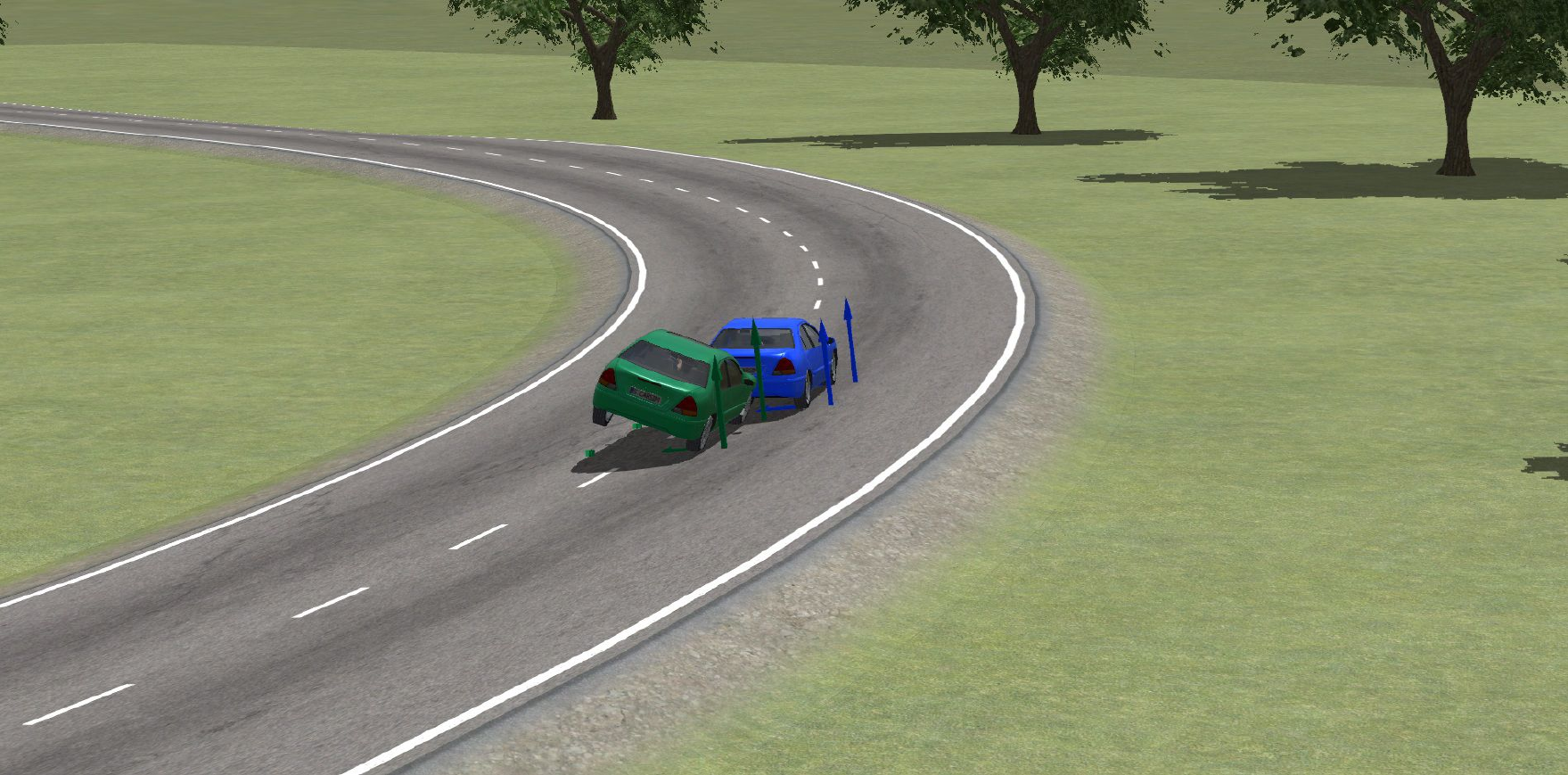}
    \caption{ A snapshot of the CarSim environment illustrating the implementation of DeePC1 and LMPC1. The image captures the moment at $t=21.7s$, where the green vehicle (LMPC1) violates the LTR constraint and begins to lose stability, while the blue vehicle (DeePC1) successfully navigates the sharp turn. }
    \label{fig:sim_result}
\end{figure*}

Initially, the RD-DeePC is tested on a sedan vehicle, demonstrating its superiority over the human driven and the LMPC. The simulation for sedan vehicle is carried out under normal road conditions with a tire-road friction coefficient of 0.85. It spans a duration of 35 seconds and includes three sharp turns, see Fig.~\ref{fig:road_geo}.
The steering wheel angle and target speed, calculated by either RD-DeePC or LMPC, are imported into the CarSim model. The steering wheel angle is directly utilized by the vehicle, while the target speed is provided as a command to the vehicle and is used by the vehicle through a low-level controller. To fully evaluate the performance of the rollover control strategy, the initial condition for the vehicle speed is $105 km/h$ and is required to navigate these turns at a speed as close to the reference speed $v_{ref} = 105 km/h$. The reference steering wheel angle is calculated from the human driver model, hence the $u_r$ in \eqref{eq:min_d_deepc} is \(u_r = [\delta_{f,ref}, v_{ref}]^T\). For both RD-DeePC and LMPC the inputs constraint sets are $u_1 = \delta_f\in [-200,200], u_2 = v_{x} \in [100, 120] =$ and the output constraint set is $ y = \text{LTR} \in [-1,+1]$. Also for both schemes, the prediction horizon is $N = 100$ and the sampling time is $T_s = 0.01s$. Finally, for both methods, $Q = 0$ in \eqref{eq:min_d_deepc} and \eqref{eq:MPC}, meaning only the input reference tracking is desired. For RD-DeePC the values for $\lambda_g$ and $\lambda_y$ are $\lambda_g = 100, \lambda_y = 10^8$ in \eqref{eq:min_d_deepc}. For LMPC, the matrices $A, B, C, D$ in \eqref{eq:lin_model} are identified from system identification using the MATLAB toolbox. 
\subsubsection{Results}
\begin{figure}[!h]
    \centering
    \includegraphics[width=0.9 \linewidth]{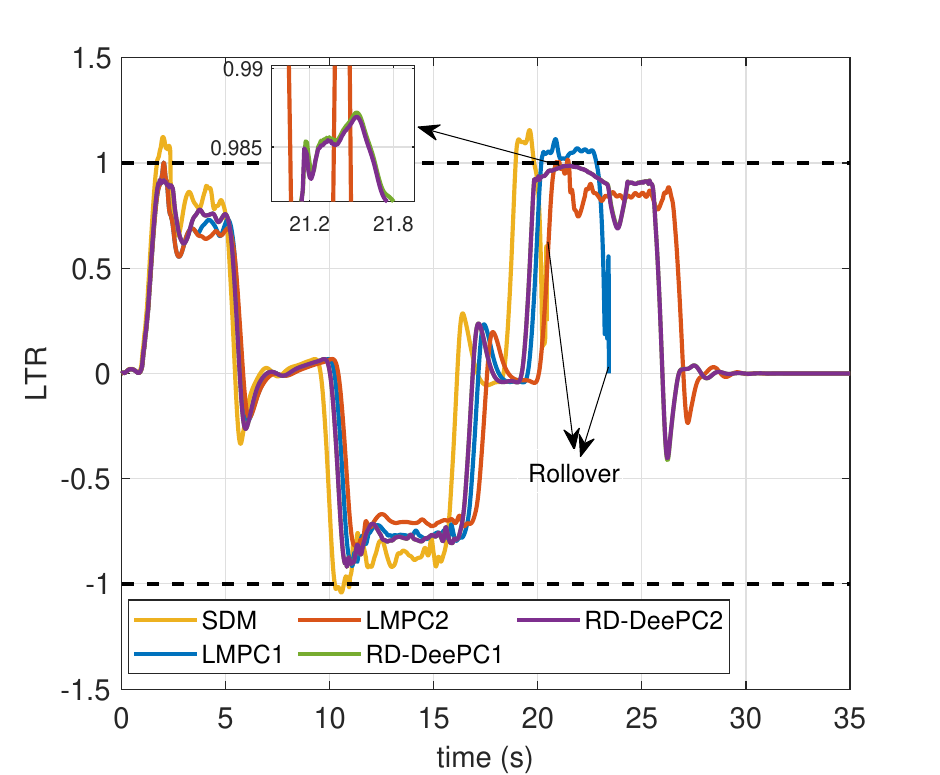}
    \caption{ Comparison of the different control strategies in terms of Load Transfer Ration (LTR) for sedan vehicle.}
    \label{fig:LTR_normal}
\end{figure}

\begin{figure}[!h]
    \centering
    \includegraphics[width=0.9 \linewidth]{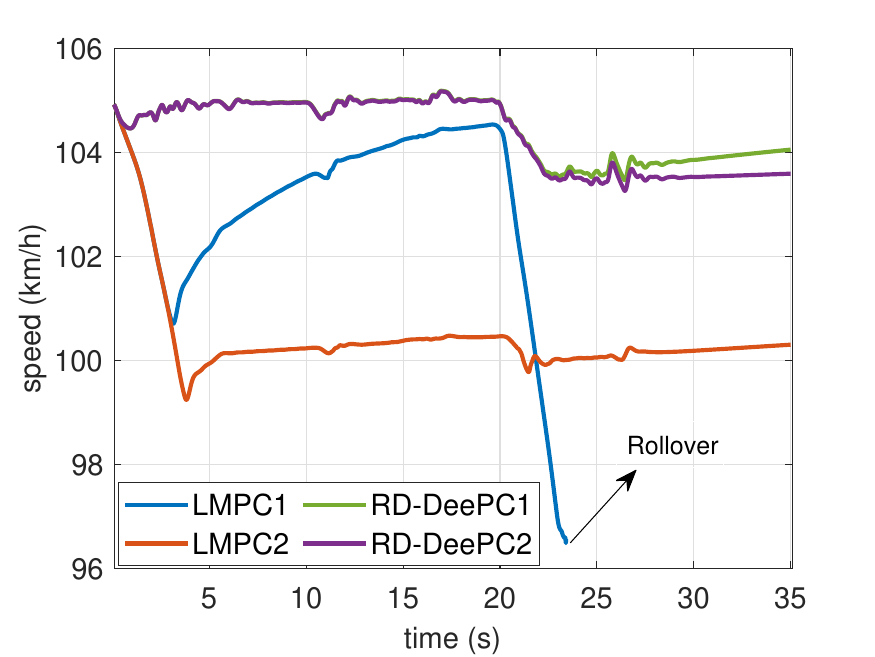}
    \caption{Comparison of the different control strategies in terms of the longitudinal speed for sedan vehicle.}
    \label{fig:ref_speed_normal}
\end{figure}

\begin{figure}[!h]
    \centering
    \includegraphics[width=0.9 \linewidth]{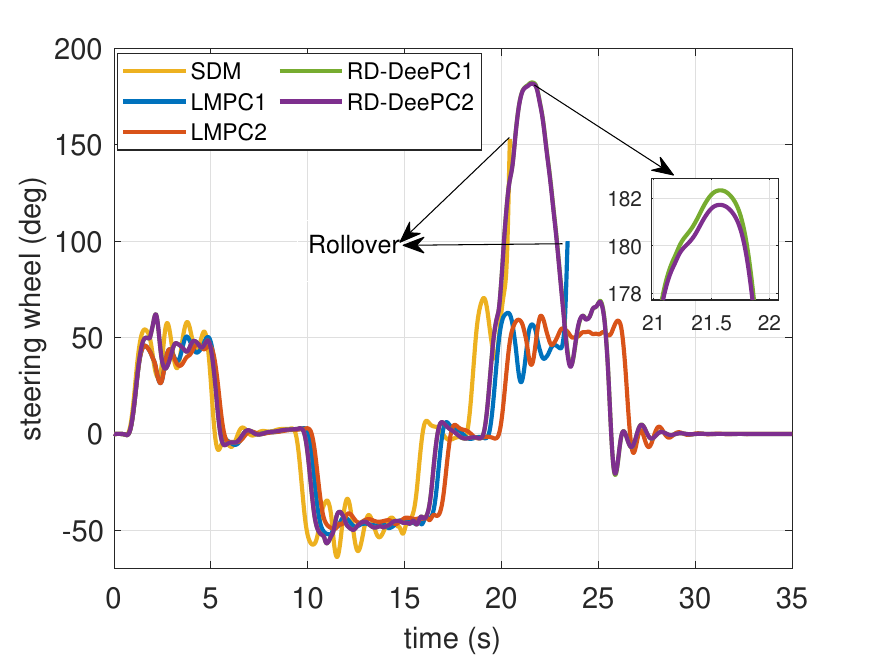}
    \caption{ Comparison of the different control strategies in terms of steering wheel for sedan vehicle}
    \label{fig:SW_normal}
\end{figure}

For the simulation of sedan vehicle, we considered two scenarios: (1)  the matrix $R$ for input reference tracking gain is $R= \mathrm{diag}(1,5\times 10^{-4})$. In this case the simulations results are denoted by RD-DeePC1 and LMPC1, and (2)  the matrix $R$ for input reference tracking gain is $R= \mathrm{diag}(1,1\times 10^{-4})$ with the corresponding RD-DeePC and MPC denoted as RD-DeePC2 and LMPC2. The rationale behind these choices is to illustrate the trade-offs between control objectives and provide a fair comparison between the LMPC and RD-DeePC methods. Although the combinations of gain selections are countless, we determined that these two configurations effectively showcase the performance of our framework.
The simulation results for the Sedan vehicle are presented in Figs.~\ref{fig:LTR_normal}-\ref{fig:SW_normal}, illustrating the output (LTR) and control inputs (i.e., the longitudinal speed and steering wheel) of the vehicle, respectively.

In the scenario (1), for LMPC1, the applied control gains successfully enabled the vehicle to navigate the first two turns but resulted in a rollover at the final turn, which happened around $t = 23$s, see Fig.~\ref{fig:LTR_normal}. Note that at this point, the LTR of for the vehicle violates the constraints and exceeds $+1$. 
For the LMPC2 control strategy, the control signals ensured safe cornering; however, LMPC2's performance was compromised, indicating that the vehicle tended to operate near to the lower speed limit to pass all turns as shown in Fig.~\ref{fig:ref_speed_normal}. This is attributed to the reduced emphasis on tracking the reference speed, which results from the specific choice of the $R$ matrix in this case. However, in this case the LTR constraint is also violated at around $t = 21s$ causing the vehicle to violate the critical safety condition, however, the vehicle did not fully lose stability and rollover did not happen. 

The human driving simulation is shown here by Simple Driver Model (SDM), an embedded proportional-integral-derivative (PID) controller in CarSim. This model generates a steering angle command to minimize the error between the vehicle's current heading or trajectory and a desired path.
The SDM strategy also led to vehicle rollover at the final turn, in addition to violating the LTR constraint several times. 

In contrast to the previous control strategies, RD-DeePC1 and RD-DeePC2 demonstrated superior performance by safely navigating the sharp turn without rollover. The vehicle speed is adjusted in conjunction with the steering wheel command to ensure the vehicle safely navigates all turns. These corrections are applied only when necessary, particularly when a high risk of rollover is detected, as occurs near the final and sharpest turn, where the adjustments to the reference signals are most significant. In RD-DeePC1, the controller places greater emphasis on target speed tracking, leading to a more aggressive effort to bring the vehicle to the reference speed in a shorter time frame after passing the last turn.
To demonstrate RD-DeePC's versatility, we next evaluate its effectiveness on different vehicle models.

\subsection{Utility truck}

\begin{figure}[h]
    \centering
    \includegraphics[width=1.1 \linewidth]{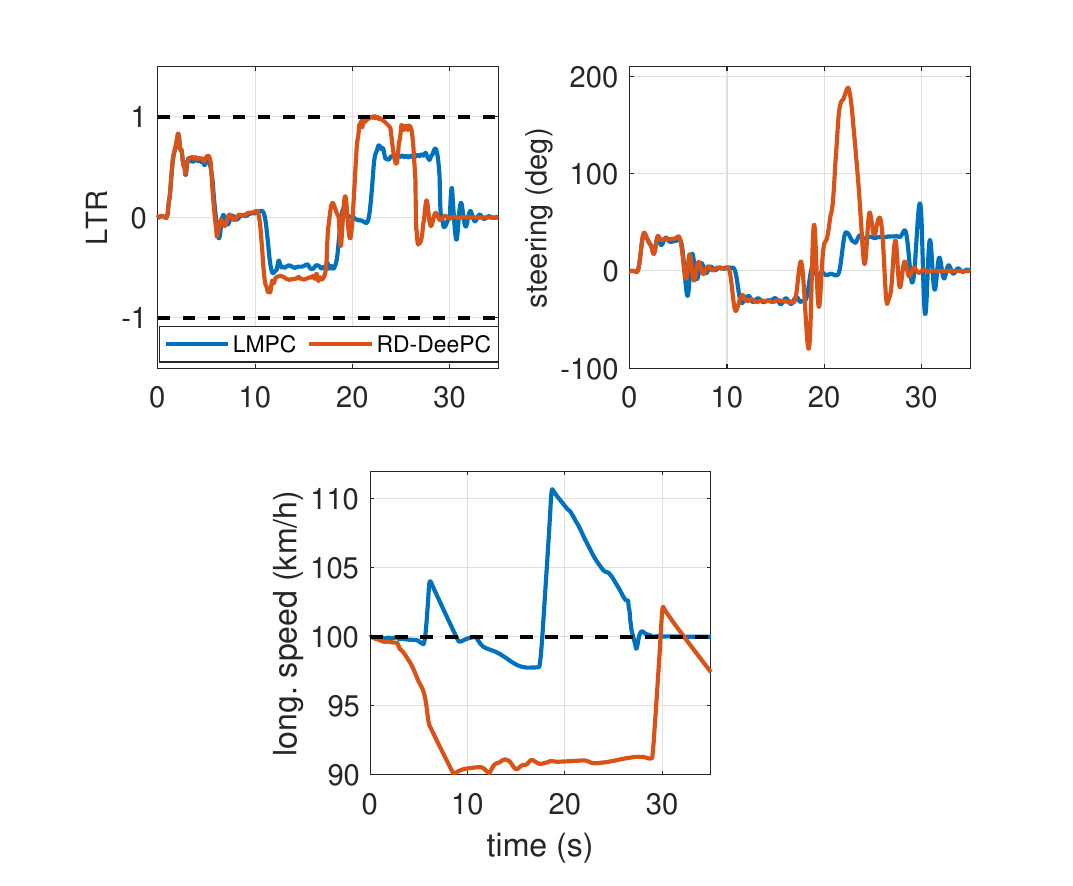}
    \caption{ Comparison between RD-DeePC and LMPC for utility truck.}
    \label{fig:truck}
\end{figure}

\subsubsection{Simulation setting}
In this subsection, the RD-DeePC strategy is evaluated on a Utility Truck, given the critical importance of addressing vehicle rollover risks for this category of vehicles. The simulation setup for both LMPC and RD-DeePC frameworks is identical. The primary distinction lies in the data collection process for the truck, with the resulting input/output data being used for the construction of the Hankel matrix and model identification specific to LMPC and RD-DeePC, respectively.  The reference speed in this case is taken as $v_{ref} =100 km/h$ and hence the constraint set for this case is changed as $u_2 = v_{x} \in [90,120]$. For LMPC, the matrices $A,B,C,D$ for the utility truck in 
\eqref{eq:lin_model} are obtained from system identification. 

\subsubsection{Results}
The results are shown in Fig.~\ref{fig:truck} for $R= diag(1,5\times 10^{-4})$ for both RD-DeePC and LMPC. As illustrated in the figure, both LMPC and RD-DeePC demonstrated the ability to successfully navigate all turns on the test course. However, there are differences in their performance characteristics. 
The LMPC approach, while ensuring safety, operated at speeds close to the minimum allowable threshold. This conservative speed profile resulted from prioritizing adherence to constraints, at the expense of performance. Such strategy can be effective in avoiding failures, but it can limit its applicability in scenarios requiring faster responses or higher efficiency.
In contrast, RD-DeePC demonstrated superior performance by completing the test without experiencing any instances of rollover or LTR violations. This suggests that RD-DeePC effectively balanced safety and speed, maintaining stability while operating at faster speeds. Its data-driven nature have contributed to a better adaptation to the test conditions, enabling it to achieve higher performance metrics without compromising safety.

\subsection{Robustness}
\begin{figure}[t]
    \centering
    \includegraphics[width=0.9 \linewidth]{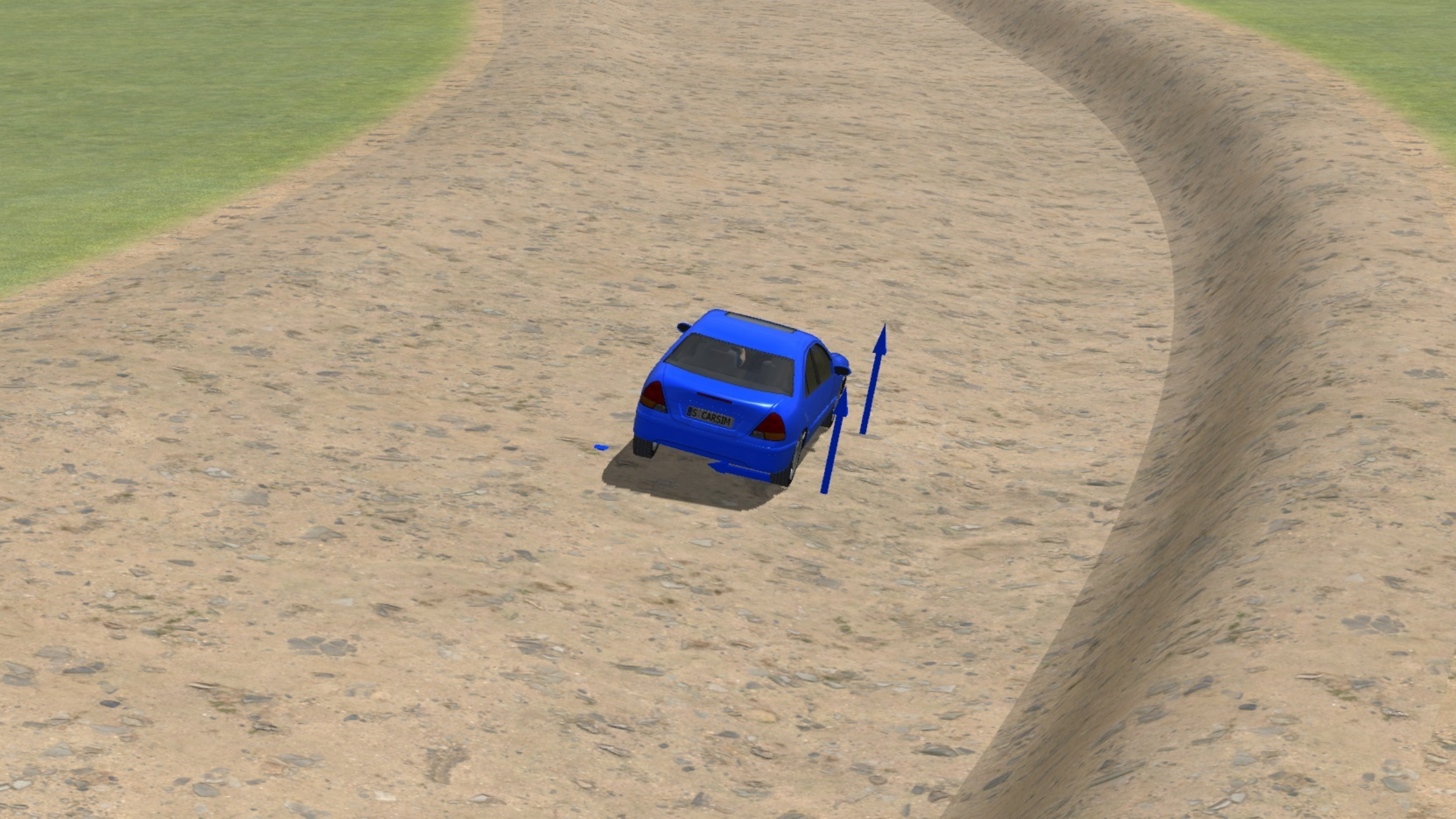}
    \caption{ Riverbed road used for robustness test.}
    \label{fig:riverbed}
\end{figure}

\begin{figure}[t]
    \centering
    \includegraphics[width=1.0 \linewidth]{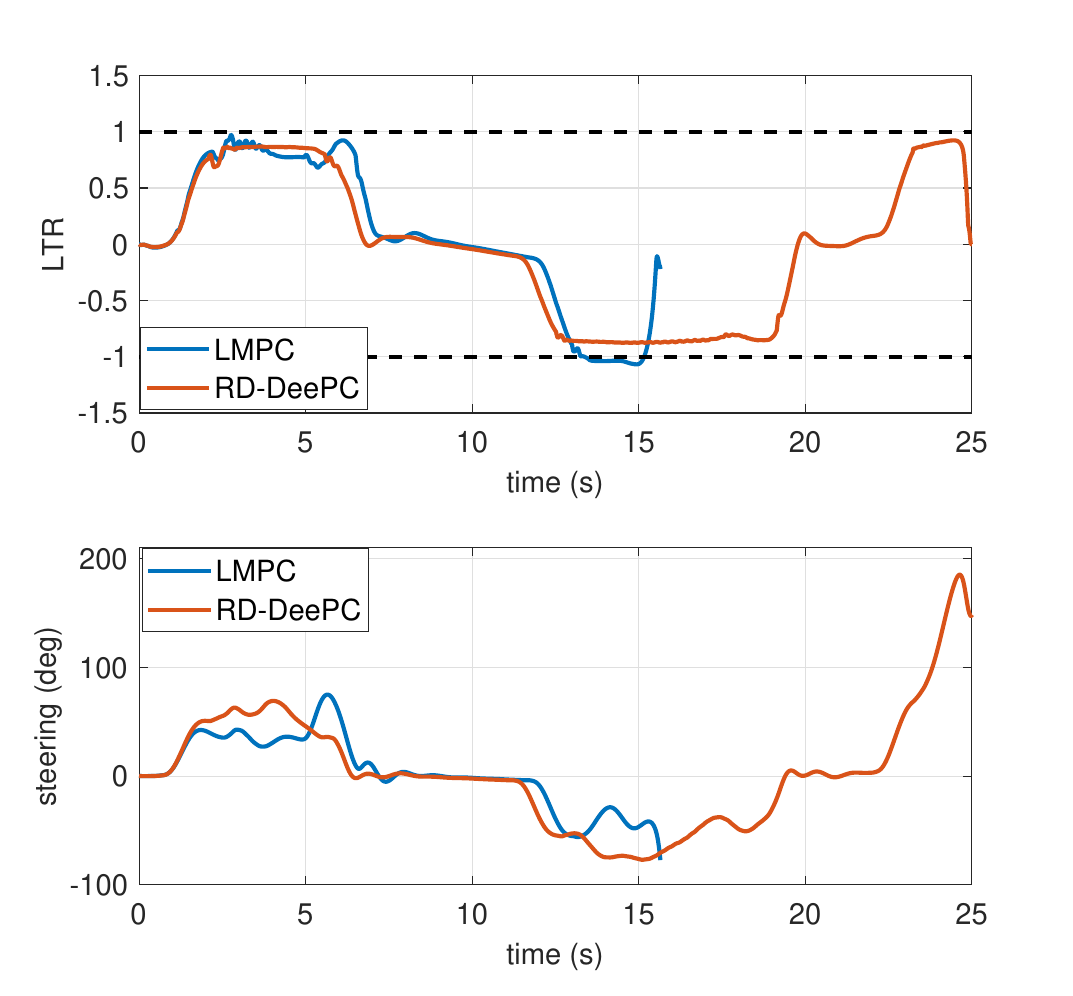}
    \caption{ Comparison of RD-DeePC and LMPC in terms of robustness to riverbed road.}
    \label{fig:robust}
\end{figure}

To evaluate the robustness of the control methodologies, a supplementary dataset containing limited input/output data from riverbed driving conditions was integrated into the existing dataset, which was originally collected under normal road conditions. This enhanced dataset was employed for model identification for the LMPC and construction of the Hankel matrix for DeePC.

The robustness evaluation was conducted by simulating a driving scenario where the road surface was altered from a normal road to a riverbed, while the geometric layout of the driving path remained identical to the scenarios previously analyzed in the study. This modification introduced more challenging and less predictable driving environment. A riverbed road includes sudden changes in elevation, such as abrupt rises and dips, and varying road bank angles, which can introduce lateral slopes. These characteristics create an uneven and unpredictable driving surface, significantly affecting vehicle stability, traction, and control dynamics compared to standard roads;
see Fig.~\ref{fig:riverbed} for a snapshot of the road condition. 

The simulation results shows a notable difference in the performance of the two control methods. RD-DeePC successfully demonstrated its robustness by adapting to the altered road conditions and avoiding vehicle rollover. This indicates that its data-driven approach effectively captured and leveraged the dynamic behavior of the vehicle under riverbed conditions. In contrast, LMPC was unable to prevent rollover, suggesting that its model-based framework was less effective in handling the abrupt change in road surface dynamics. These findings highlight the superiority of RD-DeePC in scenarios where the operating conditions deviate significantly from the nominal data used for model development.

\section{Conclusion}\label{sec:conclusion}
This paper introduced a reduced-dimension Data-Enabled Predictive Control (RD-DeePC) as a novel solution for vehicle rollover prevention. Unlike traditional control methods, RD-DeePC directly leverages raw input-output data to predict system behavior and optimize control actions, eliminating the need for explicit dynamic modeling. Simulation results using high-fidelity CarSim models for sedan and utility truck scenarios demonstrated that RD-DeePC outperforms Linear Model Predictive Control (LMPC) in preventing rollovers while ensuring better speed and maneuverability. By dynamically adjusting steering inputs and target speeds, DeePC maintained stability even during extreme driving maneuvers. The robustness and adaptability of DeePC to different vehicle types and road conditions underscore its potential as a versatile and effective control strategy for enhancing road safety. Future work may explore real-world implementation and further optimization of computational efficiency to enable deployment in resource-constrained environments.

\bibliographystyle{ieeetr}
\bibliography{main}
\balance





\end{document}